\documentclass[twocolumn,showpacs]{revtex4}
\usepackage{graphicx}
\usepackage{dcolumn}

\begin{document}


\title{Magicity and occurrence of band with enhanced $B(E2)$ in neutron-rich nuclei
$^{68}$Ni and $^{90}$Zr}

\author{K. Kaneko,$^{1}$ M. Hasegawa,$^{2}$
        T. Mizusaki,$^{3,4}$ and Y. Sun$^{5,6,7}$}

\affiliation{
$^{1}$Department of Physics, Kyushu Sangyo University, Fukuoka 813-8503, Japan \\
$^{2}$Laboratory of Physics, Fukuoka Dental College, Fukuoka 814-0193, Japan \\
$^{3}$Institute of Natural Sciences, Senshu University, Kawasaki, Kanagawa, 214-8580, Japan \\
$^{4}$Center for Nuclear Study (CNS), University of Tokyo, Wako Campus of RIKEN, Wako 351-0198, Japan \\
$^{5}$Department of Physics and Joint Institute for Nuclear Astrophysics,
 University of Notre Dame, Notre Dame, IN 46556, USA \\
$^{6}$Department of Physics, Tsinghua University, Beijing 100084,
 P. R. China \\
$^{7}${Department of Physics, Xuzhou Normal University,
 Xuzhou, Jiangsu 221009, P. R. China}
}

\date{\today}

\begin{abstract}
Experimental energy spectrum and B(E2) values in $^{68}$Ni and
$^{90}$Zr indicate a double-magic character in these neutron-rich
nuclei with $N$ or $Z=40$. The data nevertheless do not show any
pronounced irregularity in two-nucleon separation energy. To
understand the underlying physics, we carry out both shell-model
and mean-field calculations. The shell-model calculation can well
reproduce all the observations. It is understood from the
mean-field results for $^{68}$Ni that the shell gap at $N=40$
disappears due to dynamical correlations of the isovector $J=0$
pairing interaction. In $^{90}$Zr, however, such a dynamic process
with the $J=0$ pairing appears not important because of the strong
contribution of the $J>0$ interaction. We study also level schemes
in the Ni isotopes and $N=$ 50 isotones. We predict a new
band built on the $0_{2}^{+}$ state in both $^{68}$Ni
and $^{90}$Zr. The states of this band are dominated by
two-particle-two-hole excitations from the $fp$-shell to the
intruder $g_{9/2}$ orbit.
\end{abstract}


\maketitle

\section{Introduction}\label{sec1}

The study of nuclear shell effects away from the valley of
stability is one of the current topics in nuclear structure
physics. The most interesting aspects are how the well-known shell
effects, such as the occurrence of magic numbers \cite{Mayer} and
the shape-coexistence phenomenon \cite{Heyde,Wood}, manifest
themselves in exotic mass regions where nuclei have unusual
combinations of neutron and proton number. There have been
intensive discussions on the issue of weakening of shell effect in
neutron-rich nuclei. For example, the spherical $N=$ 20 shell gap
for light nuclei disappears in neutron-rich isotopes, leading to
strongly deformed ground states and large $E2$ transition
probabilities between the $2_{1}^{+}$ state and ground state
$(0_{1}^{+})$. By using the shell-model approach, it has been
demonstrated \cite{Otsuka1} that the magic number at $N=$ 20
vanishes due to the proton-neutron attraction between spin-orbit
partners of maximum $j$. On the other hand, there have been
suggestions \cite{Yamagami} that the strong deformation effects
around $^{32}$Mg are induced by dynamical correlations, such as
the neutron pairing correlations.

It has been found \cite{Broda,Grzywacz,Ishii,Sorlin} by several
experiments that the neutron-rich nucleus $^{68}$Ni $(Z=28, N=40)$
shows a double-magic character: a relatively large $2_{1}^{+}$
excitation energy and a small $B(E2,0_{1}^{+}\rightarrow
2_{1}^{+})$ value, which is comparable to the cases of
double-magic nuclei $^{16}$O, $^{40}$Ca, and $^{48}$Ca. The
nucleus $^{68}$Ni lies far from the neutron drip line, and the
neutron energy gap between the $fp$-shell and the $g_{9/2}$
intruder orbit appears to be sizeable at $N=$ 40. It was discussed
in Ref. \cite{Reinhard,Grawe} that in the early mean-field
calculations, a distinct shell gap that exists in the $N=$ 40
nucleus $^{68}$Ni disappears when quadrupole correlations are
taken into account. For $^{68}$Ni, it is remarkable that this
nucleus does not show a pronounced irregularity in two-neutron
separation energy, as expected for a typical double-magic nucleus.
It was suggested \cite{Langanke} that small
$B(E2,0_{1}^{+}\rightarrow 2_{1}^{+})$ value is not a strong
evidence for the double-magic character. We may thus conclude that
the double-magicity nature in $^{68}$Ni is still controversial and
remains an open question.

In general, shell closure leads to spherical configurations for
the ground state, while breaking of a magic shell can produce
coexisting deformed states. An important indication for the
emergent deformation is the appearance of low-lying 0$^{+}$
bands. The deformed structure occurs as a consequence
of nuclear correlations, which excite nucleons from the closed
shell to a higher shell. For example, the typical double-magic
nucleus $^{56}$Ni $(Z=N=28)$ \cite{Rudolf,Otsuka,Mizusaki} is
known to have two collective bands with large deformations
coexisting with the spherical ground band. Therefore, it is very
interesting to examine theoretically whether such collective bands
exist also in $^{68}$Ni.

Similar discussions would also apply to the neutron-rich nucleus
$^{90}$Zr, which has a closed $Z=$ 40 proton subshell and a strong
$N=$ 50 neutron shell closure. This is an interesting case to
study the persistence of the $Z=$ 40 stability. Recently, energy
levels and $B(E2)$ values in $^{90}$Zr have been measured
\cite{Garrett}, which showed a double magic character: a
relatively large $2_{1}^{+}$ excitation energy and a small
$B(E2,0_{1}^{+}\rightarrow 2_{1}^{+})$ value. However, this
nucleus does not indicate a pronounced irregularity in two-proton
separation energy. Moreover, it is known that a low-lying 0$_{2}^{+}$
state exists at $Z=$ 40 in the $N=50$ isotonic chain. Hence we can
expect to see excited bands in $^{90}$Zr but perhaps
with different structure.

In this paper, we study the magicity at $N$ or $Z=40$ and
structure of excited $0_2^+$ bands in the neutron-rich nuclei
$^{68}$Ni and $^{90}$Zr. To understand the physics in a
systematical way, we perform spherical shell-model calculations
for the Ni isotopes and $N=$ 50 isotones. Conventional shell-model
calculations in the $(1f_{7/2}, 2p_{3/2}, 1f_{5/2}, 2p_{1/2}$,
$1g_{9/2})$ shell space for $N, Z=30-36$ are not possible at
present because of the huge dimension of configuration space; we
need to restrict the model space to the $2p_{3/2}, 1f_{5/2},
2p_{1/2}$, and $1g_{9/2}$ orbitals (hereafter called the
$fpg$-shell). Of course, neutron (proton) excitations from the
$1f_{7/2}$ orbit to the $fpg$-shell cannot be neglected for
$^{68}$Ni ($^{90}$Zr) \cite{Sorlin,Grawe}. Nevertheless, after all
we shall see that the variations in $B(E2)$ in the nuclei around
$^{68}$Ni ($^{90}$Zr) can be understood in terms of valence
neutrons (protons) in this restricted model space. For the Ni
isotopes, we employ an effective interaction
starting from a realistic neutron G-matrix
interaction based on the Bonn-C $NN$ potential (called VMS
interaction) \cite{Lisetskiy}. For the $N=$ 50 isotones, we use
two types of effective interactions: the proton part of the VMS
interaction and the effective interaction of Ji and Wildenthal
(called JW interaction) \cite{Ji}.

The paper is arranged as follows. In Sections~\ref{sec2} and
\ref{sec3}, we present the numerical calculations and discuss the
results for Ni isotopes and $N=$ 50 isotones, respectively.
Conclusions are drawn in Sec.~\ref{sec4}.

\section{Ni isotopes}\label{sec2}

\subsection{Magicity in $^{68}$Ni}

\begin{figure}[t]
\includegraphics[width=8cm,height=10cm]{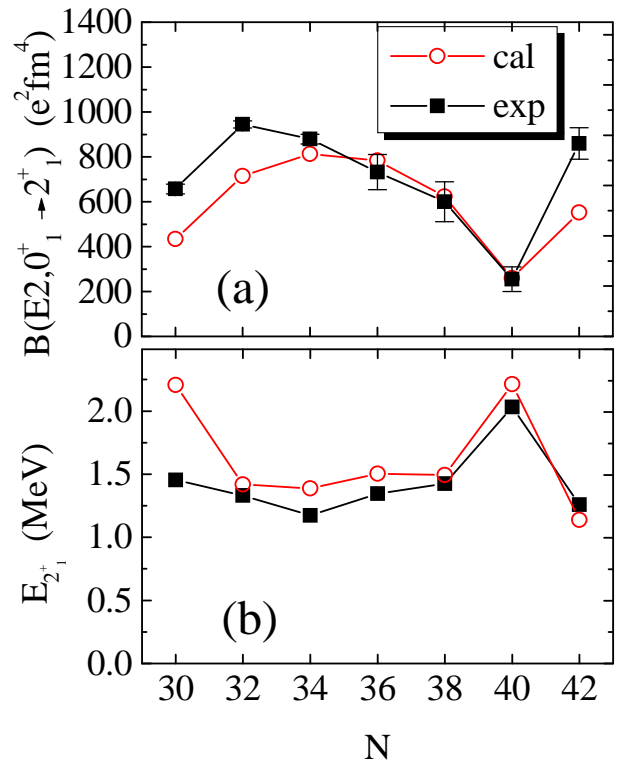}
\caption{(Color online) Comparison between the calculated and
experimental values of (a) $B(E2,0_{1}^{+}\rightarrow 2_{1}^{+})$
and (b) $E_{2_{1}^{+}}$ for the Ni isotopes. The calculated values
are denoted by open circles and the experimental data
\cite{Broda,Sorlin,Firestone,Perru} by solid squares.}
\label{fig1}
\end{figure}
\begin{figure}[t]
\includegraphics[width=8cm,height=10cm]{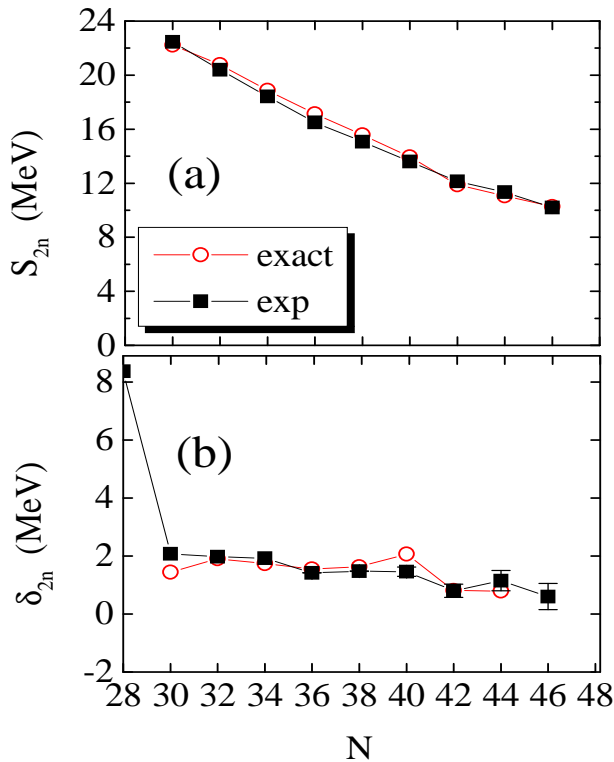}
\caption{(Color online) (a) Two-neutron separation energies and
(b) differences between the two-neutron separation energies
defined in Eq. (3). The exact shell-model results are denoted by
open circles and the experimental data \cite{Audi,Firestone} by
solid squares.}
\label{fig2}
\end{figure}
\begin{figure}[t]
\includegraphics[width=8cm,height=6cm]{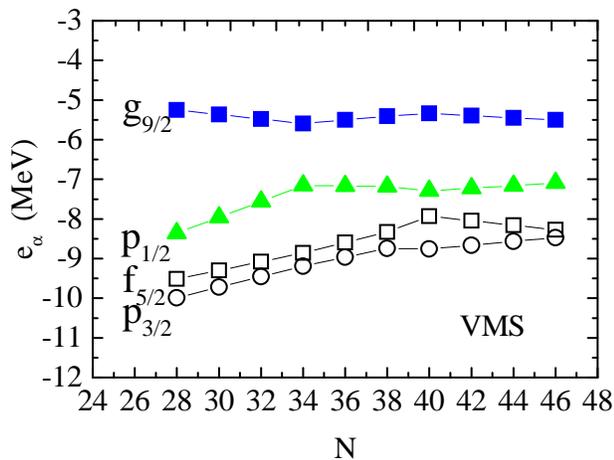}
\caption{(Color online) Spherical neutron shell structure. The HF
single-particle energy levels $e_{\alpha}$ predicted in the HF
calculations with the VMS interaction for the Ni isotopes.}
\label{fig3}
\end{figure}

Let us first review what the experiment has found for the Ni
isotopes. In Fig. 1, the experimental $B(E2,0_{1}^{+}\rightarrow
2_{1}^{+})$ value and the first excited $2^{+}$ energy are shown
as a function of neutron number $N$ for $^{58-70}$Ni. With
increasing $N$, $B(E2)$ decreases quickly and becomes the smallest
at $N=$ 40. In contrast, changes in the $2_{1}^{+}$ energy
($E_{2_{1}^{+}}$) are quite small for $^{58-66}$Ni, but
$E_{2_{1}^{+}}$ jumps to a large value at $N=$ 40. Thus, with a
pronounced large $E_{2_{1}^{+}}$ and a small
$B(E2,0_{1}^{+}\rightarrow 2_{1}^{+})$, these data seem to suggest
a subshell closure at $N=40$ in $^{68}$Ni.

We now carry out shell-model calculations for the Ni isotopes. The
shell-model Hamiltonian is written as
\begin{eqnarray}
H  & = & \sum_{\alpha}\varepsilon_{\alpha}c^{\dag}_{\alpha}c_{\alpha} +
\frac{1}{4}\sum_{\alpha\beta\gamma\delta}V_{\alpha\beta, \gamma\delta}
c^{\dag}_{\alpha}c^{\dag}_{\beta}c_{\delta}c_{\gamma},
\label{eq:1}
\end{eqnarray}
where $\varepsilon_{\alpha}$ are single-particle energies and
$V_{\alpha\beta, \gamma\delta}$ two-body matrix elements. Since
$^{56}$Ni is taken as a core, the model space is restricted to the
$fpg$-shell for neutrons and protons are assumed to be inactive.
The proton core-excitations from $^{56}$Ni are taken into account
implicitly by the effective two-body matrix elements and the
proton contributions are estimated from the KB3 calculations
\cite{Caurier} in $^{48}$Ca. The neutron effective charge is taken
as $e_{n}=$ 1.0 so as to reproduce the experimental
$B(E2,0_{1}^{+}\rightarrow 2_{1}^{+})$ of $^{68}$Ni
\cite{Lisetskiy}. We use the VMS interaction starting from a
realistic neutron G-matrix interaction based on the Bonn-C $NN$
potential.

As one can see in Fig. 1, our calculations reproduce nicely the
observed trends in $B(E2,0_{1}^{+}\rightarrow 2_{1}^{+})$ and
$E_{2_{1}^{+}}$ \cite{Lisetskiy}. In particular, a large
$2_{1}^{+}$ excitation energy and small $B(E2,0_{1}^{+}\rightarrow
2_{1}^{+})$ value at $N=$ 40 are correctly obtained. It should be
pointed out that the proton core excitations may significantly
contribute to the excitation energy and to
$B(E2,0_{1}^{+}\rightarrow 2_{1}^{+})$ in $^{58}$Ni, and thus it
is difficult to absorb these effects into the effective
interaction and the effective charges. In addition, the very
recent observation \cite{Perru} indicates a large
$B(E2,0_{1}^{+}\rightarrow 2_{1}^{+})$ value in $^{70}$Ni, which
exceeds the calculated one. Figure 2 shows two-neutron separation
energy $S_{2n}$ and the difference between two-neutron separation
energies $\delta_{2n}$, defined respectively by
\begin{eqnarray}
S_{2n}(Z,N)  & = & B(Z,N) - B(Z,N-2),
\label{eq:2}
\\
\delta_{2n}(Z,N)  & = & S_{2n}(Z,N) - S_{2n}(Z,N+2).
\label{eq:3}
\end{eqnarray}
In Eq. (\ref{eq:2}), $B(Z,N)$ is the binding energy taken as
positive values. The quantity $\delta_{2n}$ is known as the most
sensitive and direct signature for a (sub)shell closure. Our shell
model calculations reproduce well the experimental values of
$S_{2n}$ and $\delta_{2n}$. As can be seen in Fig. 2, $S_{2n}$ and
$\delta_{2n}$ are smooth functions, and in particular, do not show
any notable changes at $N=$ 40. Thus, $^{68}$Ni has a large
$E_{2_{1}^{+}}$ and a small $B(E2,0_{1}^{+}\rightarrow
2_{1}^{+})$, but no irregularity in $S_{2n}$ and no strong peak in
$\delta_{2n}$. It is therefore very interesting to further look
into the $S_{2n}$ and $\delta_{2n}$ results from the viewpoint of
the magicity in $^{68}$Ni.

Let us analyze the shell-model results in Fig. 2 using mean-field
procedures. We carry out Hartree-Fock (HF) and
Hartree-Fock-Bogolyubov (HFB) calculations using the shell-model
Hamiltonian (1). In the calculations, we impose spherical
symmetry. The HF single-particle energies are given by
\begin{eqnarray}
e_{\alpha} & = & \varepsilon_{\alpha} + \sum_{\beta =
{\rm occup}}V_{\alpha\beta, \alpha\beta},
\label{eq:4}
\end{eqnarray}
where $\sum_{\beta ={\rm occup}}$ means the summation over the
occupied states only. Figure 3 shows the HF single-particle
energies $e_{\alpha}$. The single-particle energy gap between
$g_{9/2}$ and $fp$-shell varies from 4 MeV at $N=$ 28 to 2.5 MeV
at $N=$ 40, which shows a persistence of a large shell gap at this
neutron number. As we shall discuss below, this gap in the static
single-particle picture will be washed out by dynamic
correlations.

\begin{figure}[t]
\includegraphics[width=8cm,height=10cm]{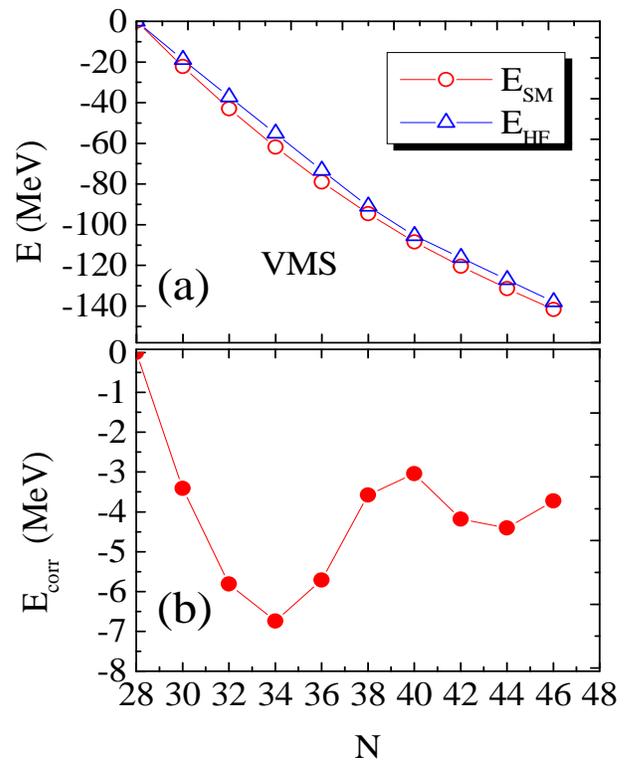}
\caption{(Color online)(a) Total energies and (b)correlation
energies in the shell-model and the HF calculations with the VMS
interaction for the Ni isotopes. Note that the absolute
correlation energies become small around $N=$ 40.}
\label{fig4}
\end{figure}

The total HF energy is expressed as
\begin{eqnarray}
E_{\rm HF} & = & \sum_{\alpha}\left( \varepsilon_{\alpha} +
\frac{1}{2}\sum_{\beta ={\rm occup}} V_{\alpha\beta, \alpha\beta}
\right).
\label{eq:5}
\end{eqnarray}
On the other hand, the HFB approximation is carried out with the
following procedure. The HFB transformation is given by
\begin{eqnarray}
a^{\dag}_{\alpha} & = & u_{\alpha}c^{\dag}_{\alpha} -
v_{\alpha}c_{\bar{\alpha}},
\label{eq:4a}
\end{eqnarray}
where $\bar{\alpha}$ is the time reversed state to $\alpha$ and
the occupation numbers $v_{\alpha}^{2}$ satisfy the following
equation
\begin{eqnarray}
v_{\alpha}^{2} & = & \frac{1}{2}
\left( 1 - \frac{\tilde{e}_{\alpha}-\lambda}{\sqrt{(\tilde{e}_{\alpha}-\lambda)^{2}+\Delta_{\alpha}^{2}}}
 \right).
\label{eq:5a}
\end{eqnarray}
Here the self-consistent mean-fields, the self-consistent pairing
gaps, and the canonical single-particle energies are respectively
defined as
\begin{eqnarray}
\Gamma_{\alpha} & = & \sum_{\beta}V_{\alpha\beta, \alpha\beta}v_{\beta}^{2}, \\
\Delta_{\alpha} & = & \sum_{\beta}V_{\alpha\bar{\alpha}, \beta\bar{\beta}}u_{\beta}v_{\beta}, \\
\tilde{e}_{\alpha} & = & \varepsilon_{\alpha} + \Gamma_{\alpha},
\label{eq:6}
\end{eqnarray}
and the total HFB energy \cite{Reinhard1} is
\begin{eqnarray}
E_{\rm HFB} & = & \sum_{\alpha}\left[ (\varepsilon_{\alpha} + \frac{1}{2}\Gamma_{\alpha})v_{\alpha}^{2}
- \frac{1}{2}\Delta_{\alpha}u_{\alpha}v_{\alpha}  \right].
\label{eq:7}
\end{eqnarray}
The neutron chemical potential $\lambda$ is determined by the
neutron number conservation
\begin{eqnarray}
\sum_{\alpha}v_{\alpha}^{2} & = & N.
\label{eq:8}
\end{eqnarray}
Eqs. (\ref{eq:5a}) and (\ref{eq:8}) are solved iteratively. In this
paper, however, we get the solutions by minimizing the total HFB
energy (\ref{eq:7}) with the neutron number conservation
(\ref{eq:8}) under the normalization condition
$u_{\alpha}^2+v_{\alpha}^2=1$.

The total shell-model energies $E_{\rm SM}$ and the HF energies
$E_{\rm HF}$ are plotted in Fig. 4(a), and the correlation
energies, defined as $E_{\rm corr}=E_{\rm SM} - E_{\rm HF}$, are
shown in Fig. 4(b). The correlation energy exhibits a
characteristic pattern where the absolute value is the largest at
$N=$ 34 but has a local minimum at $N=$ 40. The reduction in
correlation energy at $N=$ 40 would be attributed to the small
pairing gap $\Delta_{1/2}$ of the $p_{1/2}$ orbit with a small
$j$.

\begin{figure}[t]
\includegraphics[width=8cm,height=10cm]{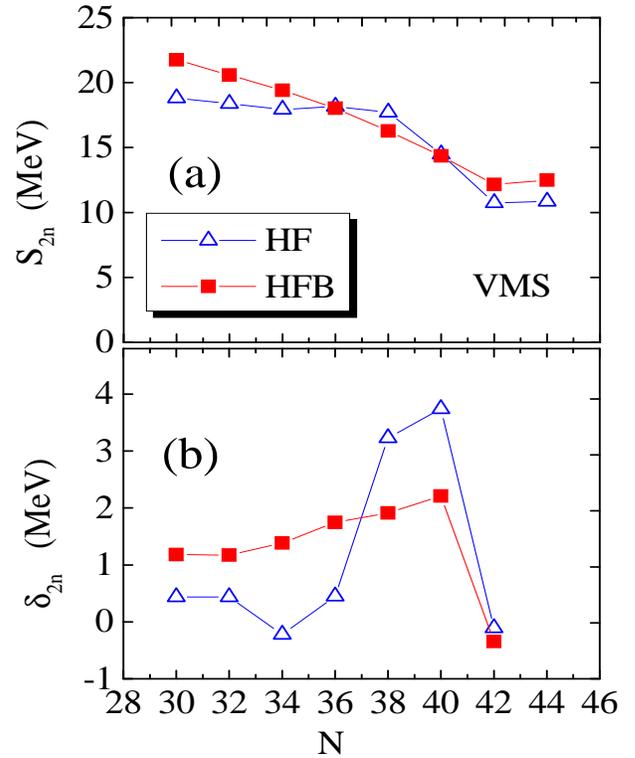}
\caption{(Color online) (a) Calculated two-neutron separation
energies and (b) differences between the two-neutron separation
energies defined by Eq. (3) in the mean-field approximation using
the VMS interaction. The HF results are denoted by open triangles
and the HFB ones by the solid squares. Note that at $N=$ 40
irregularity of $S_{2n}$  appears in the HF calculation and
disappears in the HFB calculation.}
\label{fig5}
\end{figure}

Calculations for two-neutron separation energy $S_{2n}$ and the
difference between two-neutron separation energies $\delta_{2n}$
are shown in Fig. 5. One can clearly see the irregularity in
$S_{2n}$ and a peak in $\delta_{2n}$ in the HF calculation for
$^{68}$Ni, which suggest a large energy gap and a subshell closure
at $N=40$. This result is consistent with the most of
the Skyrme HF (SHF) and relativistic mean-field
(RMF) calculations, which produced a distinct $\delta_{2n}$ peak at $N=$ 40
\cite{Reinhard,Grawe}. However, as seen in Fig. 5, the irregularity in
$S_{2n}$ and peak in $\delta_{2n}$ do not show up in the HFB
calculations when the $T=1,J=0$ pairing interaction is included.
We may therefore conclude that the $T=1,J=0$ pairing interaction
is responsible for the observed smooth behavior in $S_{2n}$ and
$\delta_{2n}$, and thus for the disappearance of a magicity
character in $^{68}$Ni. This conclusion is different from that of
the SHF and RMF calculations in which the disappearance of the
$\delta_{2n}$ peak is caused by quadrupole correlations
\cite{Reinhard,Grawe}.

\begin{figure}[t]
\includegraphics[width=8cm,height=10cm]{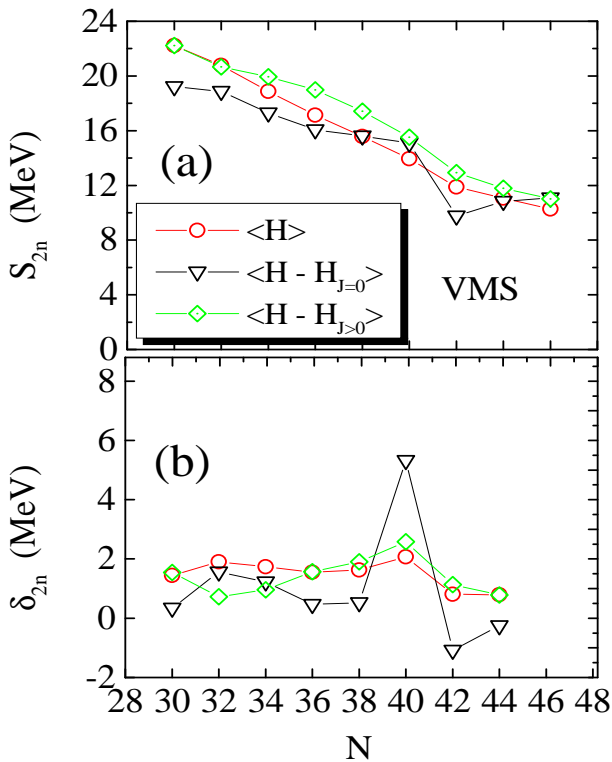}
\caption{(Color online) (a) Two-neutron separation energies and
(b) differences between the two-neutron separation energies
defined by Eq. (3) from the shell-model calculations using the VMS
interaction. The exact shell-model results are denoted by open
circles, the expectation values neglecting the $T=1,J=0$
interactions by open triangles, and the expectation values
neglecting the $J>0$ interactions by open diamonds. Note that only
the $\langle H-H_{J=0}\rangle$ result shows irregularity at $N=$
40.}
\label{fig6}
\end{figure}
\begin{figure}[t]
\includegraphics[width=8cm,height=10cm]{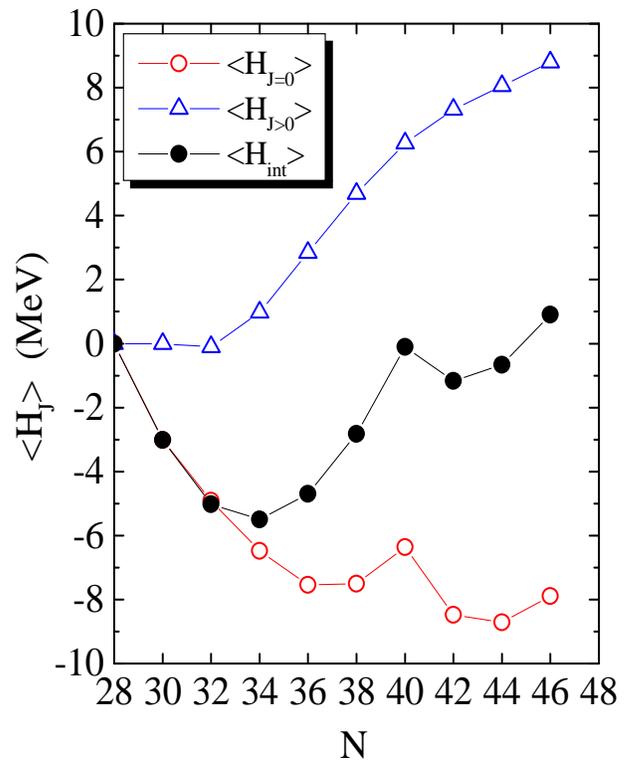}
\caption{(Color online) Expectation values of $H_{J=0}$ and
$H_{J>0}$ in Eq. (13), which are denoted by open circles and open
triangles, respectively. The total interaction energy $\langle
H_{int}\rangle$ is also depicted by solid circles. Note that
$\langle H_{J=0}\rangle$ displays a bending at $N=$ 40.}
\label{fig7}
\end{figure}

The above conclusion is reinforced by the following analysis. To
see the role of the $T=1,J=0$ pairing interaction in the shell
model calculations, we divide the two-body interaction $H_{int}$
in the total Hamiltonian (\ref{eq:1}) into two parts
\begin{eqnarray}
H_{int}  & = & H_{J=0} + H_{J>0},
\label{eq:9}
\end{eqnarray}
where $H_{J=0}$ is the $T=1,J=0$ pairing interaction and
$H_{J>0}=H-H_{J=0}$. Figure 6 compares different calculations for
$S_{2n}$ and $\delta_{2n}$. We evaluate $S_{2n}$ and $\delta_{2n}$
by using the binding energy $B(Z,N)$ calculated from the
expectation values $\langle H - H_{J=0}\rangle$ and $\langle H -
H_{J>0}\rangle$, and compare them with the results of the full
Hamiltonian. All these calculations use the same ground-state
wavefunction obtained from diagonalization of the total
Hamiltonian (\ref{eq:1}). Now the significant role of the
$T=1,J=0$ pairing interaction is clearly shown: when $H_{J=0}$ is
switched off, $S_{2n}$ exhibits irregularity and a large peak in
$\delta_{2n}$ is seen at $N=$ 40, whereas in $\langle
H-H_{J>0}\rangle$, no irregularity in $S_{2n}$ and no peak in
$\delta_{2n}$ can be seen.

In Fig. 7, we further examine the expectation values for various
Hamiltonian terms. For the quantity $\langle H_{J=0}\rangle$, one
sees that the contribution of the $T=1,J=0$ pairing causes a
bending at $N=$ 40. On the other hand, $\langle H_{J>0}\rangle$
increases monotonously with $N$. The total expectation value
$\langle H_{int}\rangle$ in Fig. 7 corresponds to the correlation
energy in Fig. 4 (b). Thus we have understood the source of the
seeming irregularity in $S_{2n}$ and the peak in $\delta_{2n}$
(see Fig. 5). The irregularity shows up in two-neutron separation
energy at $N=40$ if the $T=1,J=0$ pairing interaction is missing.
Inclusion of the $T=1,J=0$ pairing interaction washes out the
irregularities in $S_{2n}$ and $\delta_{2n}$ found in the HF
calculations, and thus explains the observations. It was inferred
from the discussion of the $g_{9/2}$ occupation number that the
erosion of the $N=40$ shell gap is attributed to the pairing
correlations \cite{Sorlin}.

\subsection{Level structure in $^{64-68}$Ni}

\begin{figure}[t]
\includegraphics[width=8cm,height=7cm]{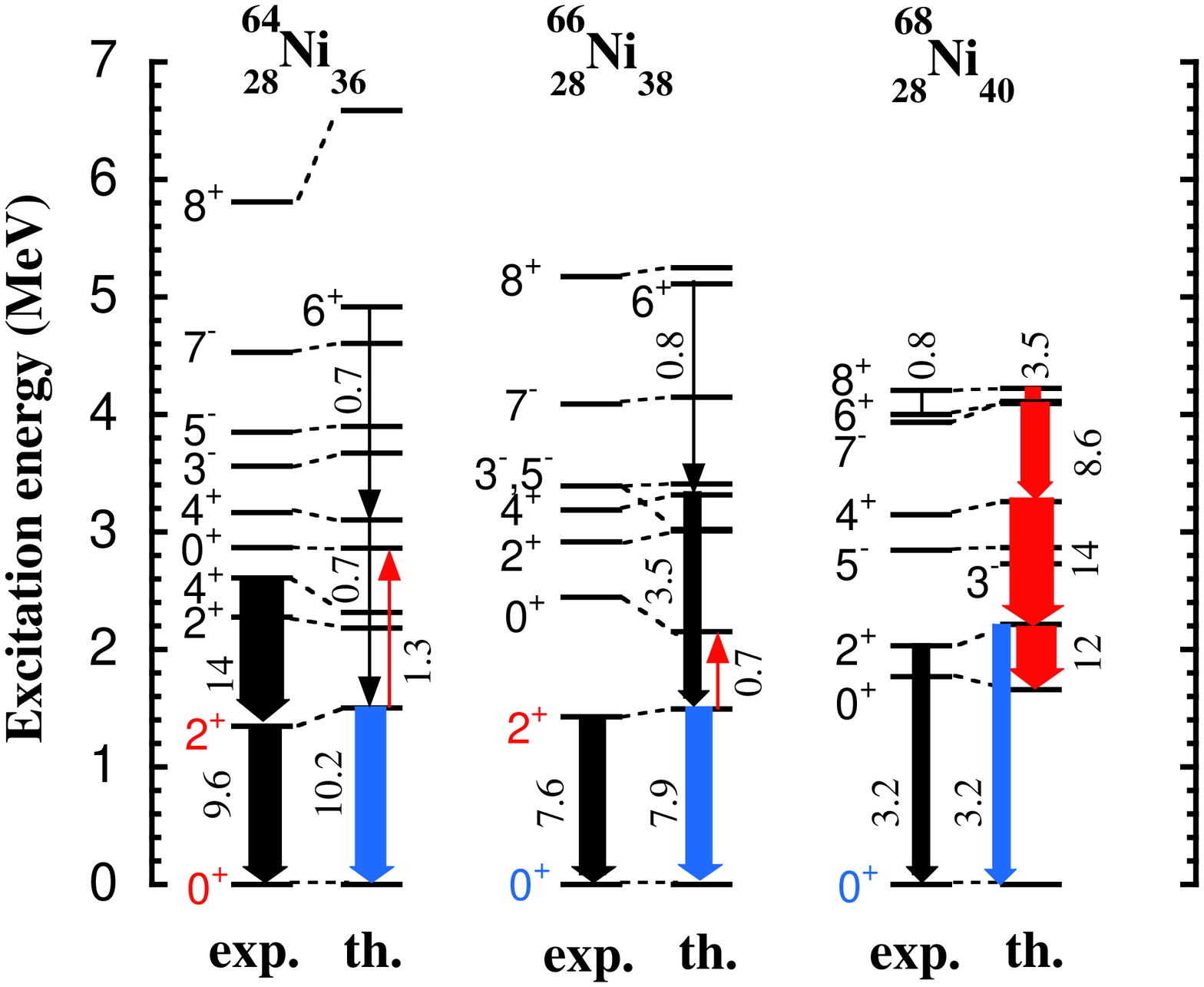}
\caption{(Color online) Comparison between the theoretical and
experimental level scheme \cite{Broda,Sorlin,Firestone} for
$^{64-68}$Ni. The widths of the arrows denote relative values of
$B(E2)$. The numbers by the arrows are the $B(E2)$ values in
Weisskopf units. }
\label{fig8}
\end{figure}
\begin{figure}[t]
\includegraphics[width=8cm,height=10cm]{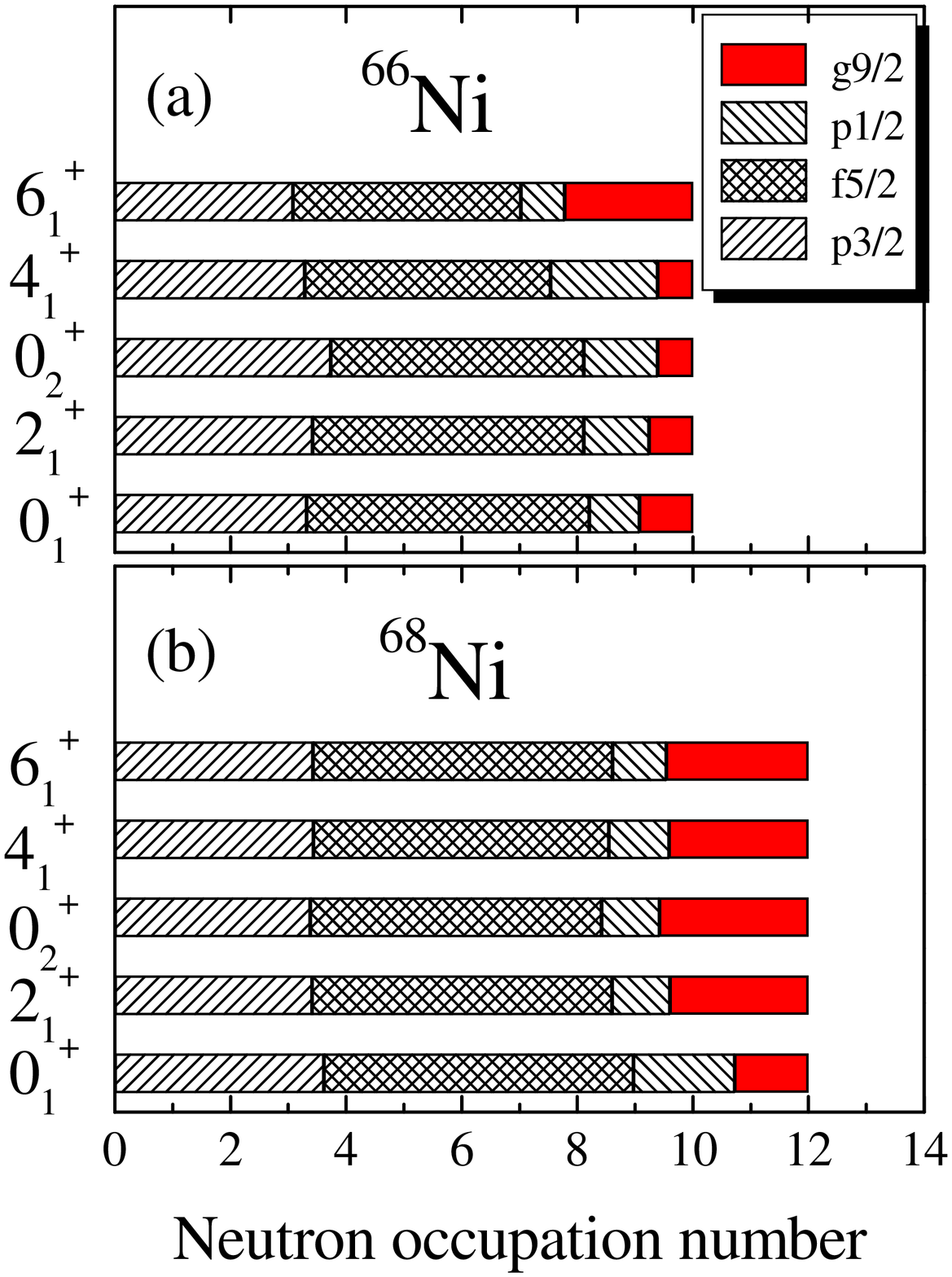}
\caption{(Color online) Neutron occupation numbers of the
$fpg$-shell orbits for the low-lying levels in (a) $^{66}$Ni and
(b) $^{68}$Ni.}
\label{fig9}
\end{figure}

In this section, we discuss the structure evolution along the
isotopic chain $^{64-68}$Ni. Figure 8 shows the experimental and
theoretical level schemes. The $B(E2)$ values have been measured
only for the first transition between the $2_{1}^{+}$ state and
the ground state \cite{Firestone}. Since in $^{64}$Ni, the
$0_{2}^{+}$, $2_{2}^{+}$, and $4_{1}^{+}$ states all lie around
2.7 MeV and their excitation energies are approximately twice the
first excited $2_{1}^{+}$ energy ($\sim$ 1.38 MeV), the level
sequence appears to be consistent with that of an harmonic
vibration. This sequence is typical for low-lying excitations in
spherical nuclei. Anharmonicity of the 2-phonon states
$(0_{2}^{+},2_{2}^{+},4_{1}^{+})$ becomes large in $^{66}$Ni, and
the harmonic pattern breaks down completely in $^{68}$Ni where the
$0_{2}^{+}$ level drops down, and appears below the $2_{1}^{+}$
level.

We carry out shell-model calculations using the VMS interaction.
The results are compared with data in Fig. 8 and the predicted
$B(E2)$ values are summarized in Table I. The calculations can
well reproduce the experimental energy levels and the
$B(E2,0_{1}^{+}\rightarrow 2_{1}^{+})$ values, and the systematic
behavior of the low-lying $0_{2}^{+}$ state is also reasonably
described. It is striking that in our results, an
excited band is formed in $^{68}$Ni based on the $0_2^+$ state.
The $E2$ transition probability $B(E2,0_{2}^{+}\rightarrow
2_{1}^{+})$ is quite small in $^{64,66}$Ni, but becomes rather
large in $^{68}$Ni. The values $B(E2,4_{1}^{+}\rightarrow
2_{1}^{+})$ and $B(E2,6_{1}^{+}\rightarrow 4_{1}^{+})$ in
$^{68}$Ni are also large. In contrast, $B(E2,0_{1}^{+}\rightarrow
2_{1}^{+})$ in $^{68}$Ni is found smaller than those in
$^{64,66}$Ni. Thus we have predicted a new band in
$^{68}$Ni, as shown in Fig. 8.

\begin{table}[b]
\caption{$B(E2)$ values for the positive-parity yrast states and
         some excited states in $^{66}$Ni and $^{68}$Ni.
         The calculated values are the shell-model results using
         the VMS interaction. Data are taken from Refs. \cite{Broda,Sorlin,Firestone}.
         }
\begin{tabular*}{85mm}{@{\extracolsep{\fill}}c|cc|cc} \hline\hline
        & \multicolumn{2}{c}{$^{66}$Ni  [$e^2$fm$^{4}$]}
        & \multicolumn{2}{c}{$^{68}$Ni  [$e^2$fm$^{4}$]}    \\ \hline
$I_i^\pi \rightarrow I_f^\pi$ & exp & cal & exp & cal \\ \hline
$2_1^+ \rightarrow 0_1^+$       & 120(20) & 125  & 53(12) & 52  \\
$4_1^+ \rightarrow 2_1^+$       &      & 56 &     & 239  \\
$6_1^+ \rightarrow 4_1^+$       &      & 13 &     & 144  \\
$8_1^+ \rightarrow 6_1^+$       &        & 79 & 26(1)  & 58  \\
$2_1^+ \rightarrow 0_2^+$     &         &  11 &      & 198  \\
$2_2^+ \rightarrow 0_1^+$     &         &  12 &      &  15    \\
$2_2^+ \rightarrow 2_1^+$     &         &  12 &      &  36    \\
$2_2^+ \rightarrow 0_2^+$     &         &  14 &      &  27    \\ \hline\hline
\end{tabular*}
\label{table1}
\end{table}

In order to see how this band is formed, we compare
neutron occupation numbers of the $p_{3/2}, f_{5/2}, p_{1/2},$ and
$g_{9/2}$ orbits in the relevant low-lying states in $^{66}$Ni and
$^{68}$Ni. As one can see in Fig. 9, except for the $6^{+}_{1}$
state, neutron occupation numbers in the low-lying states in
$^{66}$Ni are dominated by the $fp$-shell components. This is
because the Fermi energy of $^{66}$Ni lies below the $p_{1/2}$
orbit. However, neutron occupation numbers in $^{68}$Ni show very
different values \cite{Sorlin}. Except for the ground state,
occupation number of the $g_{9/2}$ orbit in all low-lying states in
$^{68}$Ni increases by more than two units. This means that two
neutrons are excited from the $fp$-shell to the $g_{9/2}$ orbit in
these states in $^{68}$Ni \cite{Grawe,Grawe1}. To see the
structure of the low-lying states more clearly, we calculate the
probability of $n$-particle-$n$-hole ($n$p-$n$h) excitations from
the $fp$-shell to the $g_{9/2}$ orbit, defined by
\begin{eqnarray}
P_{n}  & = & \frac{\langle N_{n} \rangle}{\sum_{n}\langle N_{n}
\rangle},
\label{eq:10}
\end{eqnarray}
where $N_{n}$ are the $n$p-$n$h operators from the $fp$-shell to
the $g_{9/2}$ orbit. Table II lists the probabilities of $n$p-$n$h
excitations in the relevant low-lying states of $^{66}$Ni and
$^{68}$Ni. In $^{66}$Ni, in all low-lying states except the
$6_{1}^{+}$ state, the dominant components are the 0p-0h
excitations but with considerable mixing of the 2p-2h excitations.
The 4p-4h excitations are quite small in these states. The
$6_{1}^{+}$ state in $^{66}$Ni has almost a pure 2p-2h component.
In contrast, the low-lying excited states in $^{68}$Ni show very
different structures. While the ground state has mixed 2p-2h and
0p-0h components with nearly equal probability, the low-lying
excited states have mainly the 2p-2h component with considerable
mixing with the 4p-4h excitation. A large $E_{2_{1}^{+}}$ and a
small $B(E2,0_{1}^{+}\rightarrow 2_{1}^{+})$ in $^{68}$Ni (see
Fig. 1) would be alternatively explained as follows. Once the
odd-parity $fp$ orbits are filled at $N=40$, at least two-neutrons
have to jump to the intruder $g_{9/2}$ orbit to create a
$2_{1}^{+}$ state, and therefore the energy $E_{2^{+}_{1}}$
increases \cite{Grawe,Lisetskiy,Grawe1}. The $E2$ transition between
$2_{1}^{+}$ and $0^{+}_{1}$ in $^{68}$Ni becomes small just
because the two states have different structure. Interestingly, we
indeed see from our calculation that the band is built
on the $0_{2}^{+}$ state. This happens because all the excited
states belonging to this band have a similar structure with the
2p-2h excitations.

\begin{center}
\begin{table}[t]
\caption{Probabilities of $n$ particle-hole excitations for the
low-lying states of $^{66}$Ni and $^{68}$Ni.}
\begin{tabular*}{85mm}{@{\extracolsep{\fill}}c|ccc|ccc} \hline\hline
       & \multicolumn{3}{c}{$^{66}$Ni} & \multicolumn{3}{c}{$^{68}$Ni} \\ \hline
 $I^{\pi}$        & 0p-0h     &  2p-2h    & 4p-4h     & 0p-0h    &  2p-2h     &  4p-4h  \\  \hline
 $0_{1}^{+}$     & 0.596     & 0.347     & 0.054     & 0.482    &  0.405     &  0.104  \\
 $2_{1}^{+}$      & 0.654     & 0.311     & 0.034     & 0.000    &  0.728     &  0.254  \\
 $0_{2}^{+}$      & 0.718     & 0.260     & 0.022     & 0.108    &  0.610     &  0.256  \\
 $4_{1}^{+}$      & 0.715     & 0.262     & 0.023     & 0.000    &  0.800     &  0.009  \\
 $6_{1}^{+}$      & 0.000     & 0.891     & 0.107     & 0.000    &  0.779     &  0.210  \\ \hline\hline
\end{tabular*}
\label{table2}
\end{table}
\end{center}

In order to visualize the shape of $^{68}$Ni,
we use the CHF method with the following quadratic constraint
\cite{Mizusaki}
\begin{eqnarray}
 H' & = & H + \alpha\sum_{\mu}(\langle Q_{2\mu}\rangle - q_{\mu})^{2}
            + \beta(\langle J_{x}\rangle - j_{x})^{2},
\end{eqnarray}
where $Q_{2\mu}$ and $J_{x}$ are the isoscalar quadrupole
operators and the $x$-component of angular momentum operator,
respectively. The $q_{\mu}$'s are constant parameters:
$q_{0}=\sqrt{\frac{5}{16\pi}}q{\rm cos}\gamma$, $q_{\pm
2}=\sqrt{\frac{5}{16\pi}}q{\rm sin}\gamma$, and $q_{\pm 1}=0$,
where $q$ is the isoscalar intrinsic quadrupole moment and
$\gamma$ is the triaxial angle. We set $j_{x} = \sqrt{J(J+1)}$
with $J$ the total angular momentum of the state. The parameters,
$\alpha$ and $\beta$, are taken so as to achieve a convergence for
an iteration calculation with the gradient method. Then, potential
energy surface (PES) is defined as the expectation value $\langle
H \rangle$ with respect to the CHF state for given $q$ and
$\gamma$. Figure 10 shows the contour plot of the PES in the
$q$-$\gamma$ plane for $^{68}$Ni. We find that the PES minimum
exhibits a spherical shape and an oblate softness. This is
consistent with our previous discussions on the shell-model
results, namely, a large $E_{2_{1}^{+}}$ and a small
$B(E2,0_{1}^{+}\rightarrow 2_{1}^{+})$ in $^{68}$Ni. The PES
figure in Fig. 10 is in contrast to the characteristic feature of
an oblate-prolate shape-coexistence in $^{68}$Se
\cite{Fischer,Kaneko}.

\section{$N=50$ Isotones}
\label{sec3}

\subsection{Magicity in $^{90}$Zr}

\begin{figure}[t]
\includegraphics[width=6cm,height=6cm]{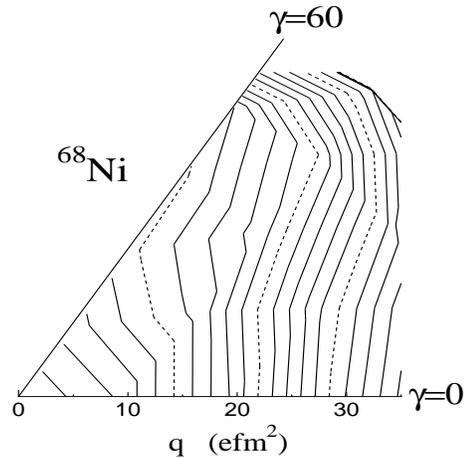}
\caption{Contour plot of PES on $q-\gamma$ plane in the CHF
calculation for $^{68}$Ni.}
\label{fig10}
\end{figure}
\begin{figure}[t]
\includegraphics[width=8cm,height=10cm]{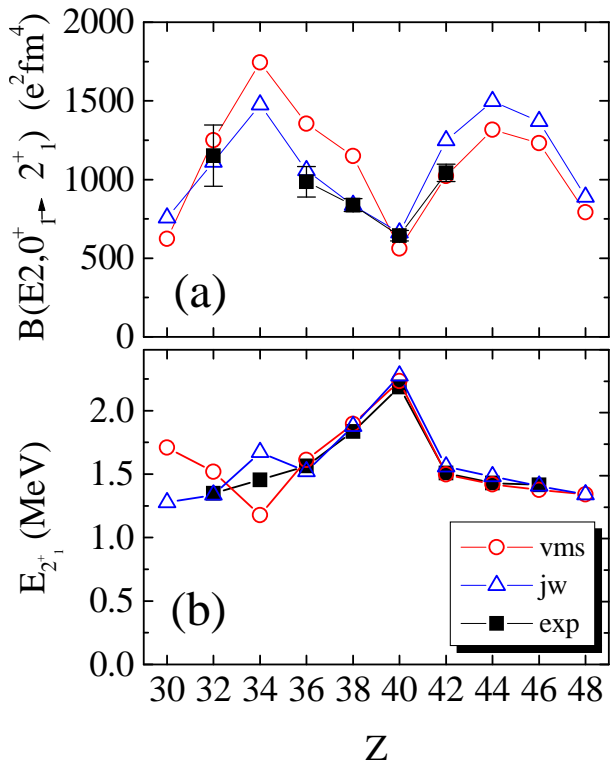}
\caption{(Color online) Comparison between the calculated and
experimental values of (a) the $B(E2,0_{1}^{+}\rightarrow
2_{1}^{+})$ and (b) $E_{2_{1}^{+}}$ for $N=$ 50 isotones. Data are
taken from Refs. \cite{Garrett,Firestone,Padilla}. The shell-model
calculations are carried out using the VMS and JW interactions.
The calculated results are denoted by open circles (triangles) for
the VMS (JW) interaction, and the experimental ones by solid
squares. }
\label{fig11}
\end{figure}
\begin{figure}[t]
\includegraphics[width=8cm,height=10cm]{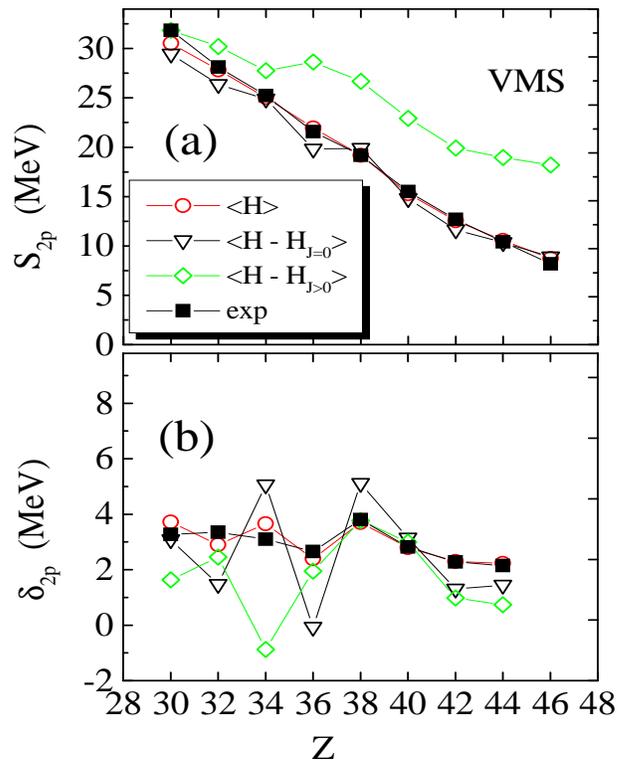}
\caption{(Color online) Two-proton separation energies in the
shell-model calculations with the VMS interaction. The exact shell
model results are denoted by open circles, the expectation values
neglecting the $T=1,J=0$ interactions by the open triangles, and
the expectation values neglecting the $J>0$ interactions by the
open diamonds. Experimental data \cite{Audi,Firestone} are denoted
by solid squares.}
\label{fig12}
\end{figure}
\begin{figure}[t]
\includegraphics[width=8cm,height=10cm]{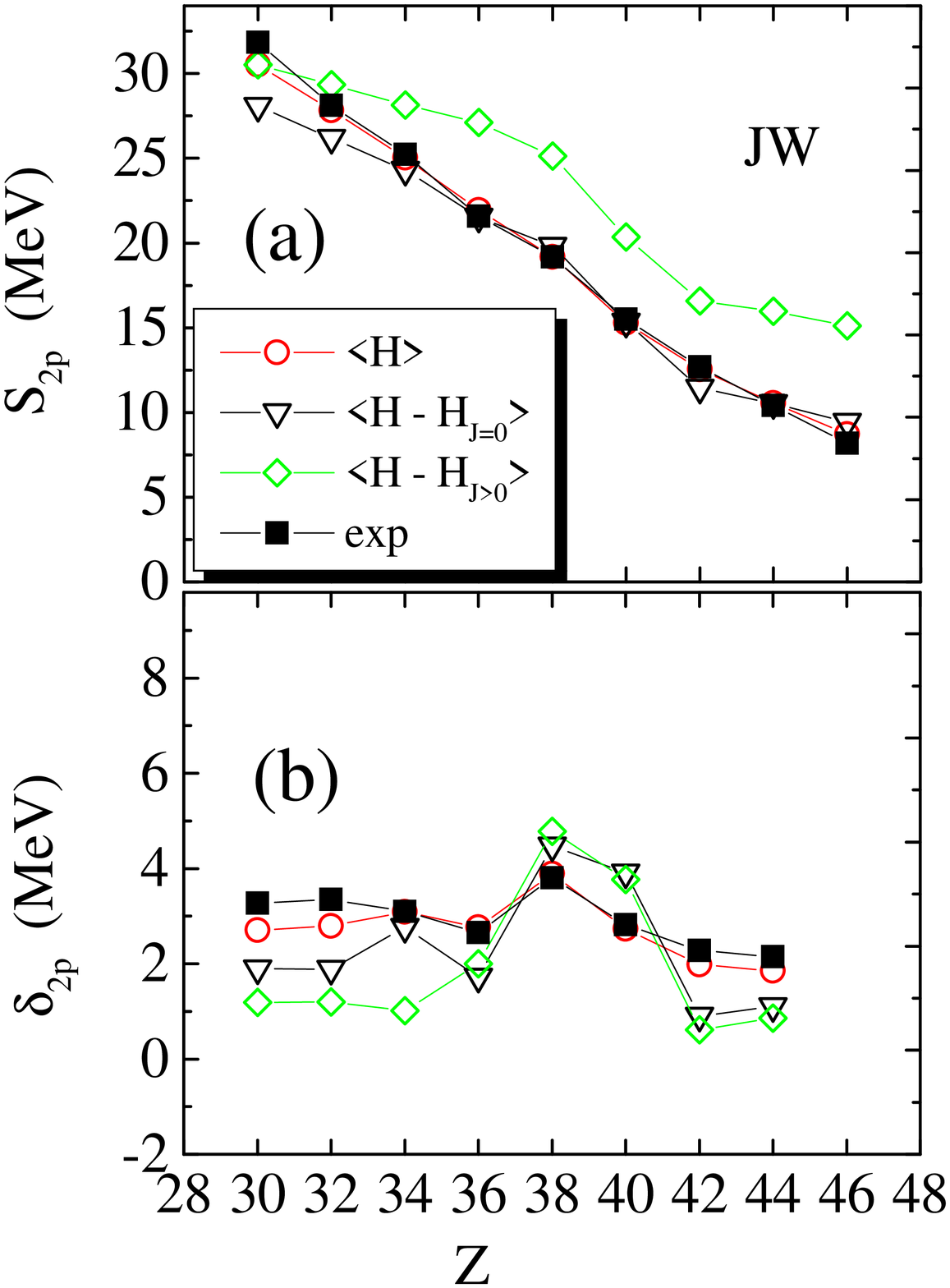}
\caption{(Color online) Same as Fig. \ref{fig12}, except that the
calculations are performed by using the JW interaction.}
\label{fig13}
\end{figure}
\begin{figure}[t]
\includegraphics[width=8cm,height=10cm]{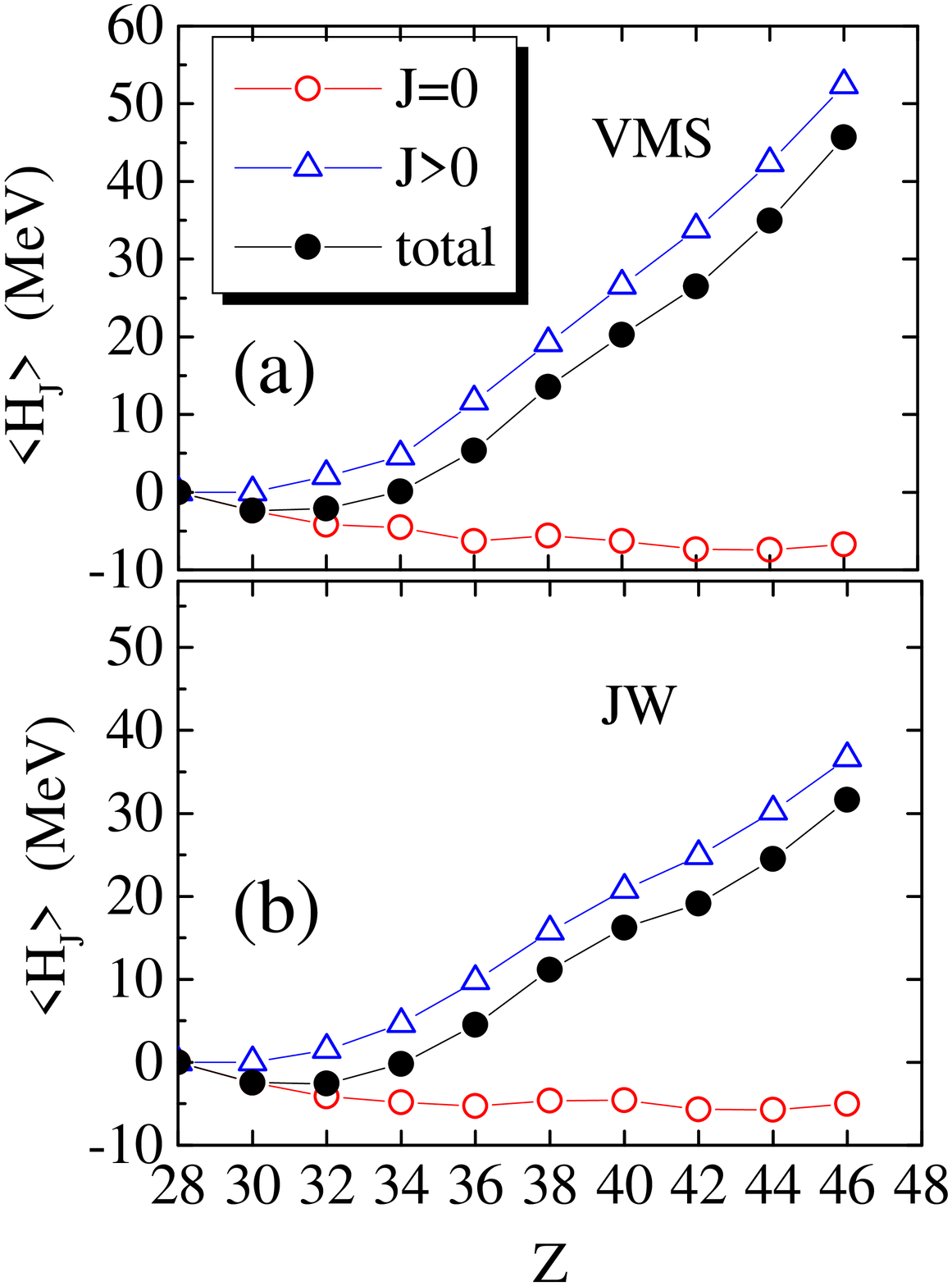}
\caption{(Color online) Expectation values of $H_{J=0}$ and
$H_{J>0}$ defined by Eq. (13) for (a) the VMS interaction and (b)
the JW interaction. $\langle H_{J=0}\rangle$ and $\langle
H_{J>0}\rangle$ are denoted by open circles and open triangles,
respectively. The total of them ($\langle H_{int}\rangle$) is also
depicted by solid circles. Note that the contributions of $\langle
H_{J>0}\rangle$ becomes large with increasing proton number. }
\label{fig14}
\end{figure}

In the previous section, we have discussed several unusual
properties found in $^{68}$Ni, which are associated with the
subshell closure at $N=40$. A related question is how neutron-rich
nuclei with $Z=40$ behave. Figure 11 shows the experimental
$B(E2,0_{1}^{+}\rightarrow 2_{1}^{+})$ and the first excited
$2_{1}^{+}$ energy as a function of proton number $Z$, for some
$N=$ 50 isotones. For both $B(E2,0_{1}^{+}\rightarrow 2_{1}^{+})$
and $E_{2^{+}_{1}}$ values in Fig. 11, we find remarkable
similarities as seen in Fig. 1: with increasing proton number $Z$,
$B(E2)$ quickly increases until $Z=$ 34 and then decreases from
$Z=$ 36 to $Z=$ 40. The first excited $2_{1}^{+}$ energy
$E_{2_{1}^{+}}$ goes up gradually and is peaked at $Z=$ 40. Again,
in terms of $B(E2,0_{1}^{+}\rightarrow 2_{1}^{+})$ and
$E_{2_{1}^{+}}$, $^{90}$Zr seems to be a double-magic nucleus. It
should be pointed out that $B(E2,4_{1}^{+}\rightarrow 2_{1}^{+})$
shows different behavior from $B(E2,0_{1}^{+}\rightarrow
2_{1}^{+})$ \cite{Lisetskiy1}. Recent lifetime measurements for
$^{96}$Pd and $^{94}$Ru corroborate the tendency of this behavior
for $N=50$ \cite{Mach}. Moreover, it has been shown recently
that the exact strengths for these transitions cannot be
reproduced in a $T=1$ model space but require neutron excitations
across the $N=50$ shell \cite{Grawe2}.

We carry out shell-model calculations for the $N=$ 50 isotones.
Since $^{78}$Ni is taken as a core, the model space for proton is
restricted to the $fpg$-shell, and the neutrons are assumed to be
inactive. The proton effective charge is taken as $e_{p}=$ 1.8 for
the VMS interaction and $e_{p}=$ 2.0 for the JW interaction so as
to reproduce the experimental $B(E2,0_{1}^{+}\rightarrow
2_{1}^{+})$ value of $^{90}$Zr \cite{Lisetskiy}. We use two types
of effective interactions: the proton part of the VMS interaction
and the JW interaction. As one can see in Fig. 11, the
calculations nicely reproduce the observed trends for both
$B(E2,0_{1}^{+}\rightarrow 2_{1}^{+})$ and $E_{2_{1}^{+}}$.

Figure 12 shows the two-proton separation energy $S_{2p}$
and the difference between two-proton separation energies
$\delta_{2p}$ for this isotonic chain, defined by
\begin{eqnarray}
S_{2p}(Z,N)  & = & B(Z,N) - B(Z-2,N),
\label{eq:11}
\\
\delta_{2p}(Z,N)  & = & S_{2p}(Z,N) - S_{2p}(Z+2,N).
\label{eq:12}
\end{eqnarray}
The experimental data do not show a signature for a subshell
closure in $^{90}$Zr, since no irregularity in $S_{2p}$ can be
seen. The shell-model calculations reproduce well the experimental
values of $S_{2p}$. In particular, the small peak in $\delta_{2p}$
at $N=38$ is well described. To understand these results, we
analyze the role of the $T=1,J=0$ pairing interaction ($H_{J=0}$)
and the other interactions ($H_{J>0}$) in the Hamiltonian, as done
in the previous section (see Eq. (\ref{eq:9})). In contrast to the
case of the Ni isotopes, Figs. \ref{fig12}(a) and Fig.
\ref{fig13}(a) indicate that, while the $T=1,J=0$ pairing
interaction scarcely contributes to $S_{2p}$, the remaining
interactions $H_{J>0}$ increases $S_{2p}$ significantly. Thus, the
$J>0$ interactions are more important for two-proton separation
energy in the $N=$ 50 isotones. The $H_{J>0}$ contribution,
however, does not produce any notable irregularity in $S_{2p}$.
For $\delta_{2p}$, we can see some differences between the VMS and
JW interactions in Figs. \ref{fig12}(b) and \ref{fig13}(b).
Moreover, the $H_{J=0}$ and $H_{J>0}$ contributions to
$\delta_{2p}$ in the VMS results are different from those in the
JW results.

Figure \ref{fig14} shows the expectation values of $H_{J=0}$,
$H_{J>0}$ and the total interaction energy $\langle
H_{int}\rangle$. Comparing Fig. \ref{fig14} with Fig. \ref{fig7},
we find that in the $N=$ 50 isotones, $\langle H_{J>0}\rangle$
increases drastically with increasing proton number, and becomes
dominant when $Z$ is large. There is no clear bending at $Z=$ 40
in either curve $\langle H_{J=0}\rangle$ and $\langle
H_{J>0}\rangle$. Thus these detailed results have explained the
trends of two-proton separation energy in Fig. \ref{fig12}(a) and
Fig. \ref{fig13}(a).

\begin{figure}[t]
\includegraphics[width=8cm,height=10cm]{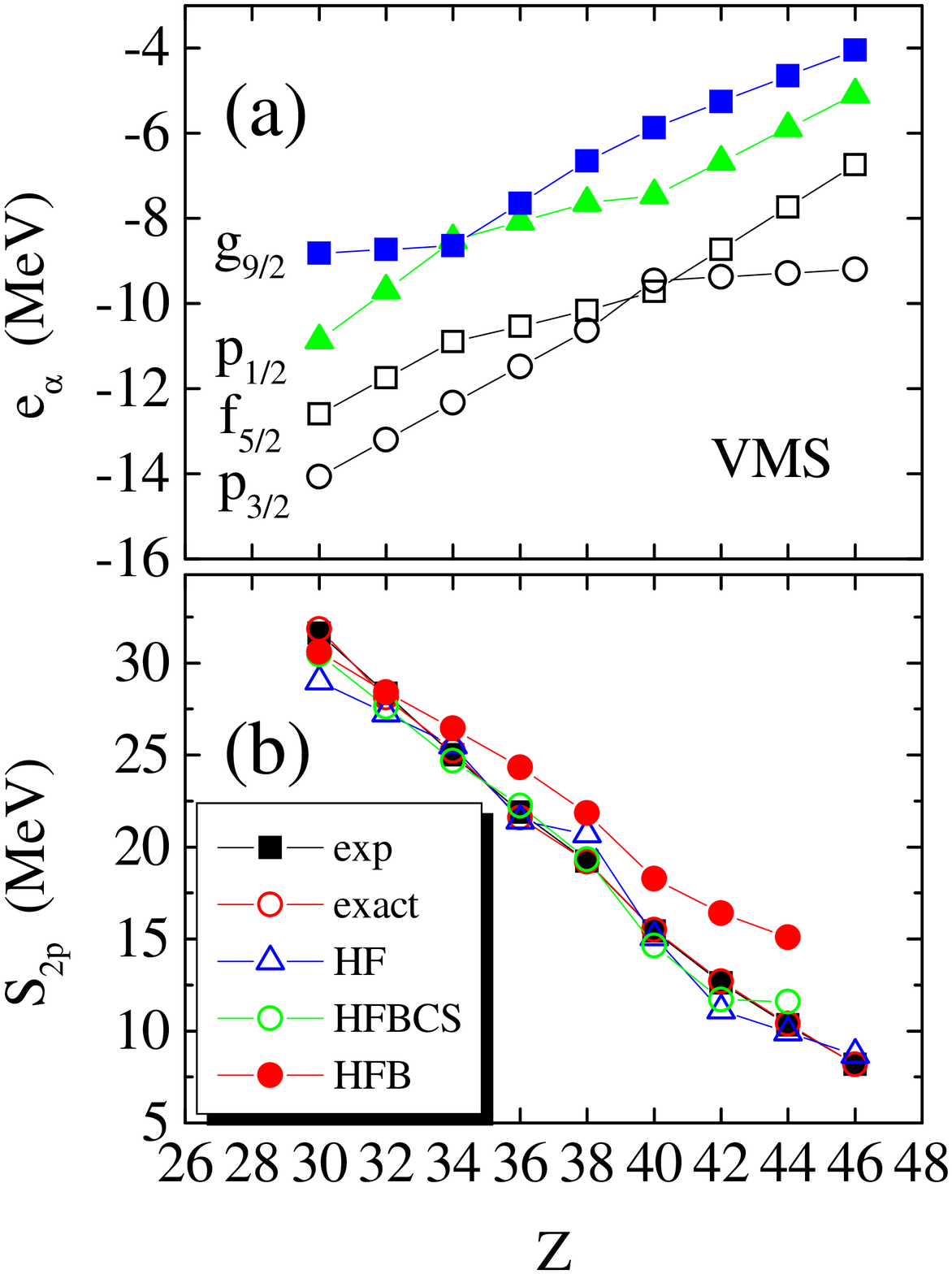}
\caption{(Color online) Spherical proton shell structure. (a) The
HF single-particle energies $e_{\alpha}$ obtained by the HF
calculations with the VMS interaction and (b) the two-proton
separation energies using the mean-field calculations for the $N=$
50 isotones. The HF results are denoted by open triangles and the
HFB ones the solid squares. }
\label{fig15}
\end{figure}
\begin{figure}[t]
\includegraphics[width=8cm,height=10cm]{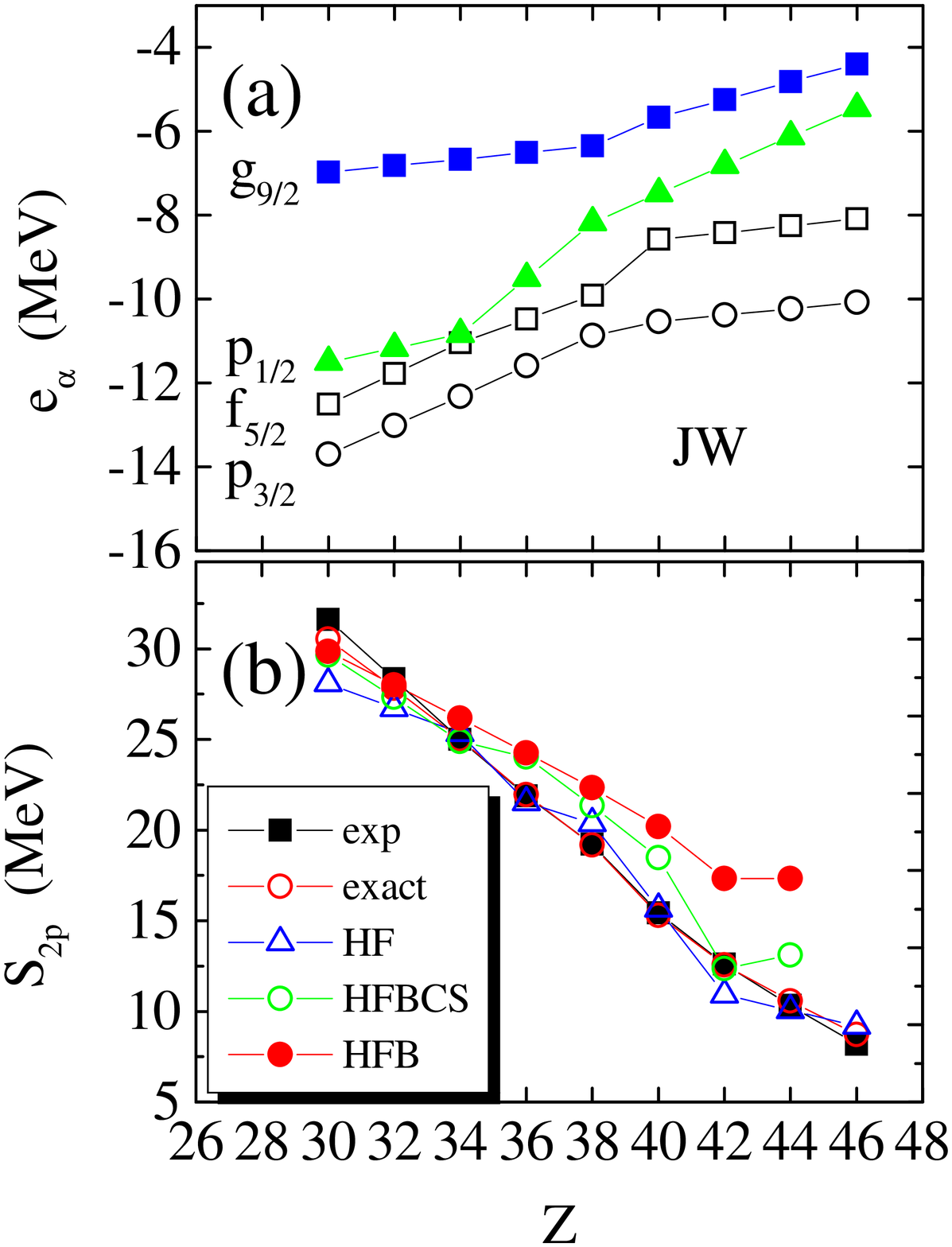}
\caption{(Color online) Same as Fig. \ref{fig15}, except that the
calculations are performed by using the JW interaction.}
\label{fig16}
\end{figure}

Let us now study the contributions from the above interactions to
HF single-particle energies $e_{\alpha}$ in the HF, HF+BCS, and
HFB treatments. We also evaluate two-proton separation energy
$S_{2p}$ within these treatments. Figures \ref{fig15} and
\ref{fig16} show respectively the results calculated with the VMS
and JW interactions. It is seen that in the VMS results shown in
Fig. \ref{fig15}(a), the single-particle energy gap between
$g_{9/2}$ and $p_{1/2}$ orbits decreases quickly with increasing
proton number. This causes a smooth variation in $S_{2p}$ as seen
in Fig. \ref{fig15}(b). All the HF type calculations do not
produce irregularity in $S_{2p}$. In the JW results in Fig.
\ref{fig16}(a), the single-particle energy gap between the
$g_{9/2}$ and $p_{1/2}$ orbits remains large up to $Z=$ 36, but
becomes small after $Z=$ 38. The Fermi energy, as found in all HF
type calculations, lies in the $fp$-shell for $Z=30-38$, and
between $g_{9/2}$ and $p_{1/2}$ for $Z=40-46$. Therefore, protons
do not encounter a large energy gap when they are excited.
Therefore, also with the JW interaction, irregularity in $S_{2p}$
is not produced (see Fig. \ref{fig16}(b)). We note that in both
Figs. \ref{fig15} and \ref{fig16}, the proton separation energies
in the HFB calculations deviate from those of the other
calculations when $Z$ is large. Similar trend is obtained in the
shell-model calculation without the $J>0$ interaction.

\subsection{Level structure in $^{86}$Kr, $^{88}$Sr, and $^{90}$Zr }

\begin{figure}[t]
\includegraphics[width=8cm,height=7cm]{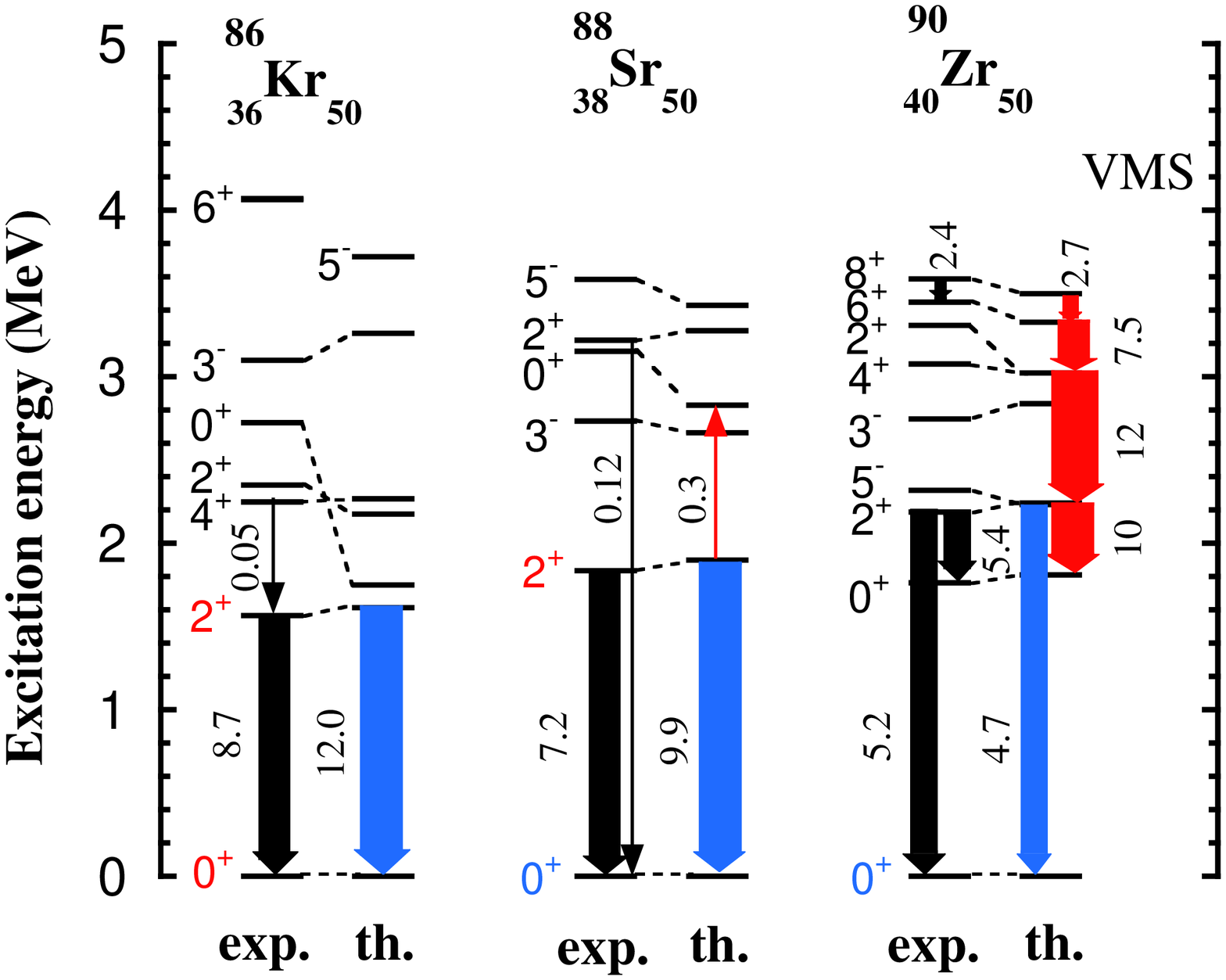}
\caption{(Color online) Comparison between the experimental and
calculated level scheme for $^{86}$Kr, $^{88}$Sr, and $^{90}$Zr.
Data are taken from Refs. \cite{Garrett,Firestone}. The shell
model calculations are carried out by using the VMS interaction.
The widths of the arrows denote relative values of B(E2). The
numbers by the arrows are the B(E2) values in Weisskopf units.}
\label{fig17}
\end{figure}
\begin{figure}[t]
\includegraphics[width=8cm,height=7cm]{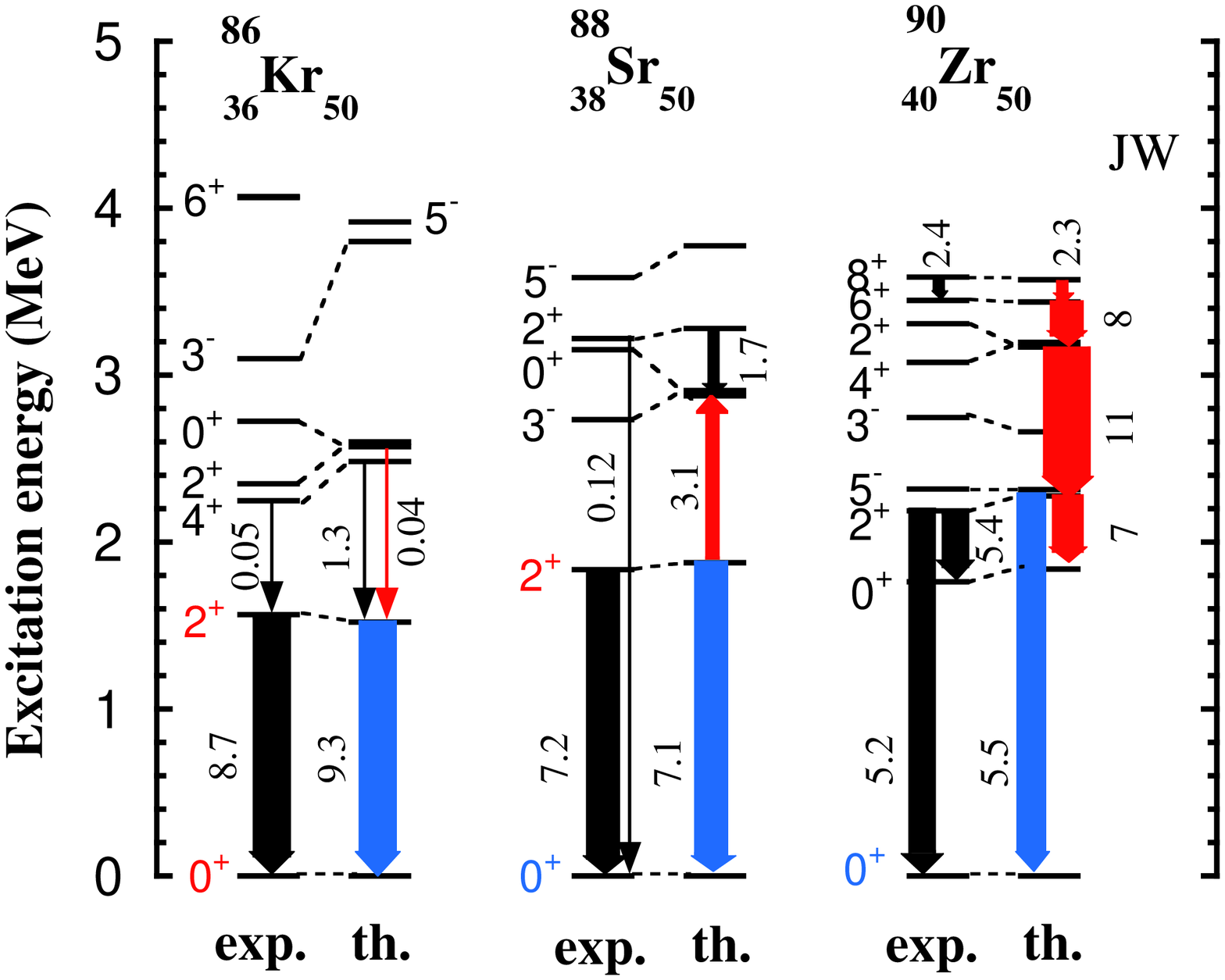}
\caption{(Color online) Same as Fig. 17, except that the
theoretical results are obtained by using the JW interaction.}
\label{fig18}
\end{figure}

Experimental level schemes for $^{86}$Kr, $^{88}$Sr, and $^{90}$Zr
are shown in Figs. \ref{fig17} and \ref{fig18}. $B(E2)$ in these
nuclei has been measured only for transitions between the
$2_{1}^{+}$ state and the ground state. Among the three isotones,
the $0_{2}^{+}$ level in $^{90}$Zr is the lowest in energy and
lies below the $2_{1}^{+}$ state.
We perform shell-model calculations using the VMS and JW
interactions, and the results are compared with data in Figs.
\ref{fig17} and \ref{fig18} and the $B(E2)$ values are summarized
in Tables III and IV.

The calculations can reproduce the experimental energy levels and
$B(E2,0_{1}^{+}\rightarrow 2_{1}^{+})$ values. In particular, the
systematical behavior of the $0_{2}^{+}$ state is well described.
It is striking that the results show again an excited
band in $^{90}$Zr based on the $0_{2}^{+}$ state. For the $E2$
transition probability $B(E2,0_{2}^{+}\rightarrow 2_{1}^{+})$,
both calculations indicate a quite small value in $^{86}$Kr and
$^{88}$Sr, but a very large one in $^{90}$Zr. Moreover,
$B(E2,4_{1}^{+}\rightarrow 2_{1}^{+})$ and
$B(E2,6_{1}^{+}\rightarrow 4_{1}^{+})$ are found large in
$^{90}$Zr. In contrast, $B(E2,0_{1}^{+}\rightarrow 2_{1}^{+})$ in
$^{90}$Zr is smaller than that in $^{86}$Kr and $^{88}$Sr. All of
these suggest strongly a new band in the $Z=40$ nucleus
$^{90}$Zr.

\begin{table}[b]
\caption{$B(E2)$ values for the positive-parity yrast states and
         some excited states in $^{88}$Sr and $^{90}$Zr.
         Data are taken from Refs. \cite{Garrett,Firestone}.
         The calculated values are the shell-model results using
         the VMS interaction.
         }
\begin{tabular*}{85mm}{@{\extracolsep{\fill}}c|cc|cc} \hline\hline
        & \multicolumn{2}{c}{$^{88}$Sr  [$e^2$fm$^{4}$]}
         & \multicolumn{2}{c}{$^{90}$Zr  [$e^2$fm$^{4}$]}    \\ \hline
$I_i^\pi \rightarrow I_f^\pi$ & exp    & cal  & exp      & cal \\ \hline
$2_1^+ \rightarrow 0_1^+$     & 167(5) & 230  & 129(4)   & 112  \\
$4_1^+ \rightarrow 2_1^+$     &        & 113  &          & 277  \\
$6_1^+ \rightarrow 4_1^+$     &        & 0.2  & $<1054$  & 180  \\
$8_1^+ \rightarrow 6_1^+$     &        &  16  &    57(4) &  65  \\
$2_1^+ \rightarrow 0_2^+$     &        & 7.3  &  124(2)  & 240  \\
$2_2^+ \rightarrow 0_1^+$     & 2.8(1) & 0.5  &          &  38  \\
$2_2^+ \rightarrow 2_1^+$     &        &  0.2 &          &  38  \\
$2_2^+ \rightarrow 0_2^+$     &        &  0.1 &          & 133  \\ \hline\hline
\end{tabular*}
\label{table3}
\end{table}
\begin{table}[b]
\caption{Same as Table III, except that the calculations are
performed by using the JW interaction.}
\begin{tabular*}{85mm}{@{\extracolsep{\fill}}c|cc|cc} \hline\hline
        & \multicolumn{2}{c}{$^{88}$Sr  [$e^2$fm$^{4}$]}
         & \multicolumn{2}{c}{$^{90}$Zr  [$e^2$fm$^{4}$]}    \\ \hline
$I_i^\pi \rightarrow I_f^\pi$ & exp & cal & exp & cal \\ \hline
$2_1^+ \rightarrow 0_1^+$     & 167(5) & 166   &    129(4)    & 133  \\
$4_1^+ \rightarrow 2_1^+$     &        &  51   &              & 264  \\
$6_1^+ \rightarrow 4_1^+$     &        &  116  &    $<1054$   & 192  \\
$8_1^+ \rightarrow 6_1^+$     &        &  226  &    57(4)     &  56  \\
$2_1^+ \rightarrow 0_2^+$     &        &  70   &    124(2)    & 173  \\
$2_2^+ \rightarrow 0_1^+$     & 2.8(1) &  117  &              &  94  \\
$2_2^+ \rightarrow 2_1^+$     &        &  23   &              &  0.05  \\
$2_2^+ \rightarrow 0_2^+$     &        &  41   &              &  60  \\ \hline\hline
\end{tabular*}
\label{table4}
\end{table}
\begin{center}
\begin{table}[t]
\caption{Probabilities of $n$p-$n$h excitations in the low-lying
states for $^{88}$Sr and $^{90}$Zr, where the VMS interaction is
used in the shell-model calculations.}
\begin{tabular*}{85mm}{@{\extracolsep{\fill}}c|ccc|ccc} \hline\hline
     & \multicolumn{3}{c}{$^{88}$Sr} & \multicolumn{3}{c}{$^{90}$Zr} \\ \hline
 $I^{\pi}$        & 0p-0h     &  2p-2h    & 4p-4h     & 0p-0h    &  2p-2h     &  4p-4h  \\  \hline
 $0_{1}^{+}$     & 0.732     & 0.244     & 0.022     & 0.425    &  0.467     &  0.100  \\
 $2_{1}^{+}$      & 0.799     & 0.190     & 0.011     & 0.000    &  0.835     &  0.158  \\
 $0_{2}^{+}$      & 0.498     & 0.433     & 0.065     & 0.300    &  0.546     &  0.141  \\
 $2_{2}^{+}$      & 0.811     & 0.180     & 0.009     & 0.000    &  0.854     &  0.141  \\
 $4_{1}^{+}$      & 0.496     & 0.464     & 0.040     & 0.000    &  0.854     &  0.141  \\
 $6_{1}^{+}$      & 0.000     & 0.920     & 0.078     & 0.000    &  0.852     &  0.142  \\ \hline\hline
\end{tabular*}
\label{table5}
\end{table}
\end{center}
\begin{center}
\begin{table}[t]
\caption{Same as Table V, except that the calculations are
performed by using the JW interaction.}
\begin{tabular*}{85mm}{@{\extracolsep{\fill}}c|ccc|ccc} \hline\hline
      & \multicolumn{3}{c}{$^{88}$Sr} & \multicolumn{3}{c}{$^{90}$Zr} \\ \hline
 $I^{\pi}$        & 0p-0h     &  2p-2h    & 4p-4h     & 0p-0h    &  2p-2h     &  4p-4h  \\  \hline
 $0_{1}^{+}$     & 0.809     & 0.184     & 0.006     & 0.452    &  0.484     &  0.062  \\
 $2_{1}^{+}$      & 0.899     & 0.100     & 0.001     & 0.000    &  0.894     &  0.104  \\
 $0_{2}^{+}$      & 0.208     & 0.729     & 0.061     & 0.356    &  0.523     &  0.115  \\
 $2_{2}^{+}$      & 0.671     & 0.320     & 0.013     & 0.000    &  0.925     &  0.074  \\
 $4_{1}^{+}$      & 0.250     & 0.714     & 0.036     & 0.000    &  0.904     &  0.094  \\
 $6_{1}^{+}$      & 0.000     & 0.967     & 0.032     & 0.000    &  0.912     &  0.087  \\ \hline\hline
\end{tabular*}
\label{table6}
\end{table}
\end{center}

To confirm the above findings, we further study the probability of
the $n$p-$n$h excitations defined by Eq. (14), in two shell-model
calculations with the VMS and JW interactions. The results for
$^{88}$Sr and $^{90}$Zr are listed in Tables V and VI,
respectively. For $^{88}$Sr with the VMS interaction (Table V),
the ground state and the $2_{1}^{+}$ state have a dominant
component of the 0p-0h excitation, and the $0_{2}^{+}$ and
$4_{1}^{+}$ states have comparable probabilities of the 0p-0h and
2p-2h excitations. Note that the $2_{1}^{+}$ state in $^{88}$Sr
can be made by 1p-1h excitations from $(f_{5/2},p_{3/2})$ to
$p_{1/2}$, which contribute to the $E2$ transitions. In the JW
results (Table VI), the 2p-2h components are dominant in the
$0_{2}^{+}$ and $4_{1}^{+}$ states. The $6_{1}^{+}$ state is
almost a pure 2p-2h excitation in both VMS and JW interactions.
One can thus expect that only the $E2$ transition $B(E2,
0_{1}^{+}\rightarrow 2_{1}^{+})$ is enhanced in $^{88}$Sr. In
$^{90}$Zr, on the other hand, the ground state has the 0p-0h and
2p-2h components with nearly equal probability and the dominant
components in the $2_{1}^{+}$, $0_{2}^{+}$, and $6_{1}^{+}$ states
are the 2p-2h excitation mixed with the 0p-0h component. Similar
results are found in both calculations. From this analysis, we can
understand that an excited band is formed on the
$0_{2}^{+}$ state in $^{90}$Zr because the $0_{2}^{+}$,
$2_{1}^{+}$, $4_{1}^{+}$, and $6_{1}^{+}$ states all have a
similar structure with a large component of the 2p-2h excitations.

\section{Conclusions}\label{sec4}

We have studied the magicity of $N$ or $Z=40$ and the level
schemes for the neutron-rich nuclei $^{68}$Ni and $^{90}$Zr by
means of the shell-model and the mean-field approximations. For
both nuclei with either $N=40$ or $Z=40$, their two-nucleon
separation energies do not show any irregularity along the
respective isotopic or isotonic chain, in spite of the apparent
double-magic feature shown with a comparatively large $2_{1}^{+}$
excitation energy and a small $B(E2,0_{1}^{+}\rightarrow
2_{1}^{+})$ value. The reason why the separation energy does not
exhibit irregularity has been found different for the Ni isotopes
and the $N=50$ isotones. From the shell-model calculations using
the VMS and JW interactions, we have suggested that the $T=1,J=0$
pairing interaction is responsible for the absence of any
irregularity in separation energy in $^{68}$Ni. The irregularity
appears in the HF treatment but disappears in the HFB treatment.
This indicates that the shell gap at $N=40$ disappears due to
dynamical correlations of the isovector $J=0$ pairing interaction.
In the case of $^{90}$Zr, however, irregularity in two-proton
separation energy does not appear in the HF calculations. For the
$N=$ 50 isotopes, the $J>0$ interactions contribute significantly
to the two-proton separation energy.

We have also studied level schemes for $^{68}$Ni and $^{90}$Zr. We
have predicted an excited band built on the $0_{2}^{+}$
state in both nuclei. The dominant component of this
band has been determined as the 2p-2h excitations from the
$fp$-shell to the intruder $g_{9/2}$ orbit. The structure of the
excited states of this band is quite different from that of the
ground state. This happens because the opposite signs of parity
between the $g_{9/2}$ orbit and the $fp$-shell do not allow 1p-1h
excitations \cite{Grawe1}. The first excited $2_{1}^{+}$ state in
$^{68}$Ni and $^{90}$Zr lies higher, and
$B(E2,0_{1}^{+}\rightarrow 2_{1}^{+})$ is relatively weak. The
difference in parity between the $fp$ and the $g_{9/2}$ subshells
leads to a small probability of quadrupole excitations across $N=$
40, and the large energy gain due to pairing correlations in the
$g_{9/2}$ subshell is responsible for the high $2^{+}$ energy in
$^{68}$Ni.



\begin{thebibliography} {99}

\bibitem{Mayer} M. G. Mayer, Phys. Rev. {\bf 75}, 1947 (1949); O. Haxel, J. H. Jensen,
and H. E. Suess, Phys. Rev. {\bf 75}, 1766 (1949).

\bibitem{Heyde} K. Heyde, P. Van Isacker, M. Waroquier, J. L. Wood, and R. M. Meyer,
Phys. Rep. {\bf 102}, 291 (1983).

\bibitem{Wood} J. L. Wood, K. Heyde, W. Nazarewicz, M. Huyse, and P. Van Duppen,
Phys. Rep. {\bf 215}, 101 (1992).

\bibitem{Otsuka1} T. Otsuka, R. Fujimoto, Y. Utsuno, B. A. Brown, M. Honma, and
T. Mizusaki, Phys. Rev. Lett. {\bf 87}, 082502 (2001).

\bibitem{Yamagami} M. Yamagami and N. Van Giai, Phys. Rev. C {\bf 69}, 034301 (2004).

\bibitem{Broda} R. Broda, {\it et al.}, Phys. Rev. Lett. {\bf 74}, 868 (1995).

\bibitem{Grzywacz} R. Grzywacz, {\it et al.}, Phys. Rev. Lett. {\bf 81}, 766 (1998).

\bibitem{Ishii} T. Ishii, {\it et al.}, Eur. Phys. J. A {\bf 13}, 15 (2002).

\bibitem{Sorlin} O. Sorlin, {\it et al.}, Phys. Rev. Lett. {\bf 88}, 092501 (2002).

\bibitem{Reinhard} P. G. Reinhard, M. Bender, T. Buervenich, C. Reiss, J. Maruhn, and
                  W. Greiner, REKIN Review, {\bf 26}, 23 (2000).

\bibitem{Grawe} H. Grawe and M. Lewitowicz, Nucl. Phys. {\bf A693}, 116 (2001).

\bibitem{Langanke} K. Langanke, J. Terasaki, F. Nowacki, D. J. Dean, and W. Nazarewicz,
                   Phys. Rev. C {\bf 67}, 044314 (2003).

\bibitem{Rudolf} D. Rudolph, {\it et al.}, Phys. Rev. Lett. {\bf 82}, 3763 (1999).

\bibitem{Otsuka} T. Otsuka, M. Honma, and T. Mizusaki, Phys. Rev. Lett. {\bf 81}, 1588 (1998).

\bibitem{Mizusaki} T. Mizusaki, T. Otsuka, Y. Utsuno, M. Honma, and T. Sebe,
                 Phys. Rev. C {\bf 59}, R1846 (1999).

\bibitem{Garrett} P. E. Garrett, {\it et al.}, Phys. Rev. C {\bf 68}, 024312 (2003).

\bibitem{Lisetskiy} A. F. Lisetskiy, B. A. Brown, M. Horoi, and H. Grawe,
                    Phys. Rev. C {\bf 70}, 044314 (2004).

\bibitem{Ji} X. Ji and B. H. Wildenthal, Phys. Rev. C {\bf 37}, 1256 (1988).

\bibitem{Firestone} R. B. Firestone and V. S. Shirley, {\it Table of Isotopes},
                    8th ed. (Wiley-Interscience, Nwe York, 1996).

\bibitem{Perru} O. Perru {\it et al.}, Phys. Rev. Lett. {\bf 96}, 232501 (2006).

\bibitem{Caurier} E. Caurier, A.P. Zuker, A. Poves, G. Mart\'{\i}nez-Pinedo,
                  Phys. Rev. C \textbf{50} (1994) 225;
                  A. Poves, A.P. Zuker, Phys. Rep. 70 (1981) 235.

\bibitem{Audi} G. Audi and A. H. Wapstra, Nucl, Phys. {\bf A595}, 409 (1995).

\bibitem{Reinhard1} P.-G Reinhard, D. J. Dean, W. Nazarewicz, J. Dobaczewski, J. A. Maruhn,
                   and M. R. Strayer, Phys. Rev. C {\bf 60}, 014316 (1999).

\bibitem{Grawe1} H. Grawe {\it et al.}, Proc. TOURS 2000, AIP Conf. Proc., Vol. {\bf 561},
                 2001, p287.

\bibitem{Fischer} S. M. Fischer, D. P. Balamuth, P. A. Hausladen, C. J. Lister,
M. P. Carpenter, D. Seweryniak, and J. Schwartz, Phys. Rev. Lett.
{\bf 84}, 4064 (2000).

\bibitem{Kaneko} K. Kaneko, M. Hasegawa and T. Mizusaki, Phys. Rev. C {\bf 70},
051301(R)(2004).

\bibitem{Padilla} E. Padilla-Rodal {\it et al.}, Phys. Rev. Lett. {\bf 94}, 122501 (2005).

\bibitem{Lisetskiy1} A. Lisetskiy {\it et al.}, Eur. Phys. J. {\bf A25}, 95 (2005).

\bibitem{Mach} H. Mach {\it et al.}, Proc. Int. Symposium {\it A New Era of Nuclear
               Structure Physics}, Niigata, Japan 2003, eds. Y. Suzuki, S. Ohya, M Matsuo,
               T. Ohtubo, World Scientific, Singapore, 2004, p277.

\bibitem{Grawe2} H. Grawe, A. Blazhev, M. Gorska, R. Grzywacz, H. Mach, and I. Mukha,
                 Eur. Phys. J. {\bf A27}, 257 (2006).

\end{thebibliography}
\end{document}